\documentclass[sigconf]{acmart}

\makeatletter
\def\@ACM@checkaffil{
    \if@ACM@instpresent\else
    \ClassWarningNoLine{\@classname}{No institution present for an affiliation}%
    \fi
    \if@ACM@citypresent\else
    \ClassWarningNoLine{\@classname}{No city present for an affiliation}%
    \fi
    \if@ACM@countrypresent\else
        \ClassWarningNoLine{\@classname}{No country present for an affiliation}%
    \fi
}
\makeatother

\AtBeginDocument{%
  }

\setcopyright{acmlicensed}
\copyrightyear{2024}
\acmYear{2024}
\acmDOI{10.1145/3649329.3656259}
\acmConference[DAC '24]{Design Automation Conference}{June 23--27,
  2024}{San Francisco, CA, USA}
\acmBooktitle{Proceedings of the 61st Design Automation Conference,
 June 23--27, 2024, San Francisco, CA, USA}
\acmPrice{15.00}
\acmISBN{978-1-4503-XXXX-X/18/06}

\usepackage[linesnumbered,ruled,vlined]{algorithm2e}
\usepackage{multirow}
\usepackage{pifont}
\usepackage{xcolor}
\usepackage{soul}
\usepackage{amssymb}

\newcommand{\squishlist}{
   \begin{list}{$\bullet$}
    { \setlength{\itemsep}{0pt}      \setlength{\parsep}{0pt}
      \setlength{\topsep}{3pt}       \setlength{\partopsep}{0pt}
      \setlength{\listparindent}{-2pt}
      \setlength{\itemindent}{-5pt}
      \setlength{\leftmargin}{1em} \setlength{\labelwidth}{0em}
      \setlength{\labelsep}{0.5em} } }

\newcommand{\squishend}{
    \end{list}  }

\setuldepth{cat}

\newcommand{\note}[1]{{\color{magenta}$\square$}}

\begin{document}

\title{GDR-HGNN: A Heterogeneous Graph Neural Networks Accelerator Frontend with Graph Decoupling and Recoupling}

\author{Runzhen Xue$^{1, 2}$, Mingyu Yan$^{1, 2, 3, *}$, Dengke Han$^{1, 2}$,\\
Yihan Teng$^{1, 2}$, Zhimin Tang$^{1, 2}$, Xiaochun Ye$^{1, 2}$, Dongrui Fan$^{1, 2}$}
\affiliation{%
  \institution{
    $^{1}$State Key Lab of Processors, Institute of Computing Technology, Chinese Academy of Sciences; \\
    $^{2}$University of Chinese Academy of Sciences; $^{3}$Shanghai Innovation Center for Processor Technologies;}
}
\email{{xuerunzhen21s, yanmingyu, handengke21s, tengyihan21s, tang, yexiaochun, fandr}@ict.ac.cn}

\renewcommand{\thefootnote}{\fnsymbol{footnote}}
\renewcommand{\shortauthors}{Xue et al.}

\begin{abstract}

Heterogeneous Graph Neural Networks (HGNNs) have broadened the applicability of graph representation learning to heterogeneous graphs. However, the irregular memory access pattern of HGNNs leads to the buffer thrashing issue in HGNN accelerators.

In this work, we identify an opportunity to address buffer thrashing in HGNN acceleration through an analysis of the topology of heterogeneous graphs. To harvest this opportunity, we propose a graph restructuring method and map it into a hardware frontend named GDR-HGNN. 
GDR-HGNN dynamically restructures the graph on the fly to enhance data locality for HGNN accelerators.
Experimental results demonstrate that, with the assistance of GDR-HGNN, a leading HGNN accelerator achieves an average speedup of 14.6$\times$ and 1.78$\times$ compared to the state-of-the-art software framework running on A100 GPU and itself, respectively.

\end{abstract}


\maketitle

\footnotetext[1]{Corresponding author is Mungyu Yan (\textit{yanmingyu@ict.ac.cn}).}

\section{Introduction}

The earlier success of graph neural networks (GNNs) has predominantly focused on homogeneous graphs (HomoGs), namely graphs with a singular type of vertices and edges.
However, many real-world datasets in complex systems are more aptly represented as heterogeneous graphs (HetGs)~\cite{yasunaga2019scisummnet, GCN-RL}.
HetGs encompass multiple categories of entities and relations, characterized by diverse types of vertices and edges, respectively. 
In contrast to HomoGs, HetGs not only encapsulate structural information but also feature-rich semantic information~\cite{HG_survey}.


Heterogeneous graph neural networks (HGNNs) are proposed to capture information in HetGs. They have reportedly demonstrated state-of-the-art (SOTA) performance in various crucial applications, including recommendation systems~\cite{li2022disentangled}, medical analysis~\cite{luo2021imas}. 
The execution of the most prevalent HGNNs can be primarily divided into four stages, namely semantic graph build (SGB) that partitions the original HetG into semantic graphs, feature projection (FP) that transforms each vertex's feature vector in semantic graphs using a multi-layer perceptron (MLP), neighbor aggregation (NA) that aggregates features from neighbors for each vertex in semantic graphs, and semantic fusion (SF) that fuses NA results across different semantic graphs for each vertex~\cite{understand_HGNN}.


Due to the unique workflow outlined above, current hardware, such as GPUs and GNN accelerators, faces challenges in efficiently executing HGNNs.
GPUs, for instance, struggle with efficiently handling irregular memory accesses stemming from the graph-topology-dependent program behavior in the NA stage~\cite{HyGCN, HiHGNN}. On the other hand, GNN accelerators tailored their hardware to GNNs, such as HyGCN~\cite{HyGCN}, lack the HGNN-oriented scheduling and executing units to process the unique workflow of HGNNs~\cite{HiHGNN}.




Recent work~\cite{HiHGNN} has proposed an HGNN accelerator, HiHGNN. This work designs a multi-lane architecture to harness parallelism between semantic graphs. 
Furthermore, it strategically schedules the execution order of semantic graphs based on their similarity to exploit data reusability.
Compared to the SOTA solution running on A100 GPU, HiHGNN achieves an 8.3$\times$ speedup~\cite{HiHGNN}.
However, the efficiency of HGNNs' acceleration is still hindered by the buffer thrashing issue. This issue, characterized by a high rate of swapping between the on-chip buffer and DRAM, is caused by the irregular memory access pattern.
Our evaluations indicate that this issue results in a substantial number of redundant accesses to DRAM, leading to a significant degradation in performance.


In this work, we initiate our exploration by scrutinizing the topology of semantic graphs, highlighting their general bipartite nature~\cite{OGB}. This observation unveils an opportunity to tackle buffer thrashing in HGNN acceleration.
To seize this opportunity, we propose a hardware frontend for restructuring semantic graphs, intended for integration into existing HGNN accelerators to mitigate buffer thrashing. Our proposed graph restructuring method involves both graph decoupling and graph recoupling.
Graph decoupling aims to separate the original semantic graph into a set of edges that do not share common vertices. Subsequently, graph recoupling utilizes the outcomes of graph decoupling to identify a vertex group, ensuring that every edge in the original semantic graph shares at least one vertex within this vertex group.
Ultimately, the semantic graph undergoes restructuring, resulting in a series of subgraphs, each characterized by a robust community structure defined by this group of vertices.

To summarize, we list our contributions as follows:
\squishlist
\item We quantitatively analyze the buffer thrashing issue in HGNN acceleration. Additionally, we identify an opportunity to address this issue through an in-deep observation of the HetG topology.

\item We propose a graph restructuring method to alleviate the buffer thrashing issue and reduce unnecessary memory accesses, capitalizing on the identified opportunity.

\item We intricately map the proposed method into a frontend hardware, named GDR-HGNN, which can be seamlessly integrated into the current HGNN accelerator to restructure graphs on the fly for enhanced data locality, thereby reducing DRAM accesses and improving performance.


\item We conduct a comprehensive evaluation of three HGNN models and three HetG datasets. Experimental results show that, with the assistance of GDR-HGNN, a SOTA HGNN accelerator achieves an average speedup of 14.6$\times$ and 1.78$\times$, and reduces DRAM access by 91.3\% and 42.9\%, compared to the SOTA software framework running on A100 GPU and itself, respectively.

\squishend


\section{Background}\label{sec:background}


\begin{table}[!t]
\centering
\caption{Notations and Corresponding Explanations.}
\label{tb:notation}
\vspace{-10pt}
\resizebox{0.45\textwidth}{!}{
\tabcolsep=0.5pt
\begin{tabular}{cc|cc}
\toprule
\textbf{Notation}                & \textbf{Explanation}                           & \textbf{Notation}                  & \textbf{Explanation}       \\ \midrule
$G$                     & heterogeneous graph                   & $V$                       & vertex set \\
$V_{src}(V_{dst})$      & source (dest) vertex set & $E$                     & edge set    \\ 
$u,\,v$                   & vertex & e ($e_{u,v}$)           & edge (from $u$ to $v$)   \\
$N(v)$ & neighbor vertex set of $v$  & $\mathcal{T}^v$           & vertex type set \\
$\mathcal{T}^e$         & edge type set         & $G^{\mathcal{P}}$         & semantic graph \\
$G_{s}^{\mathcal{P}}$     & subgraphs restructured by $G^{\mathcal{P}}$  & $\mathcal{R}$    & relation    \\
\bottomrule
\vspace{-15pt}
\end{tabular}}
\end{table}

\noindent \textbf{HetGs.}
A graph can be defined as $G=(V, E,\mathcal{T}^v,\mathcal{T}^e)$, where $V$ is the vertex set, $E$ is the edge set, $\mathcal{T}^v$ is the vertex type set and $\mathcal{T}^e$ is the edge type set. A graph is HetG when $|\mathcal{T}^v|+|\mathcal{T}^e|>2$. 
Table~\ref{tb:HetGs} lists some typical HetG datasets. Each edge type is termed as a relation $\mathcal{R} \in \mathcal{T}^e$, while an edge $e_{u,v}\in E$ starts from the source vertex $u$ and ends at the target vertex $v$. 
For example, the relation A $\rightarrow$ M in the IMDB dataset means that an actor A acts in a movie M. 

\begin{table}[!h]
\vspace{-5pt}
\centering
\footnotesize
\caption{Information of HetG Datasets.} \label{tb:HetGs}
\vspace{-10pt}
\renewcommand\arraystretch{0.6}
\setlength\tabcolsep{2pt}%
\resizebox{0.45\textwidth}{!}{
\begin{tabular}{cccc}
    \toprule
    \textbf{Dataset} &
      \makebox[0.1\textwidth][c]{\textbf{\#Vertex}} &
      \makebox[0.1\textwidth][c]{\textbf{\#Feature}} &
      \makebox[0.1\textwidth][c]{\textbf{Relations}} \\ \midrule
    \multirow{4}{*}{IMDB}   & movie (M): 4932       & M: 3489   & \multirow{4}{*}{\begin{tabular}[c]{@{}c@{}}A $\rightarrow$ M M $\rightarrow$ A\\ K $\rightarrow$ M M $\rightarrow$ K\\ D $\rightarrow$ M M $\rightarrow$ D\end{tabular}}\\
                            & director (D): 2393    & D: 3341   &          \\
                            & actor (A): 6124       & A: 3341   &          \\
                            & keyword (K): 7971     & K: ---    &          \\ \midrule
    \multirow{4}{*}{ACM}    & paper (P): 3025       & P: 1902   & \multirow{4}{*}{\begin{tabular}[c]{@{}c@{}}T $\rightarrow$ P P $\rightarrow$ T\\ S $\rightarrow$ P P $\rightarrow$ S\\ P $\rightarrow$ P -P $\rightarrow$ P \\ A $\rightarrow$ P P $\rightarrow$ A\end{tabular}}\\
                            & author (A): 5959      & A: 1902   &          \\
                            & subject (S): 56     & S: 1902   &          \\
                            & term (T): 1902        & T: ---    &          \\ \midrule
    \multirow{4}{*}{DBLP}   & author (A): 4057      & A: 334    & \multirow{4}{*}{\begin{tabular}[c]{@{}c@{}}A $\rightarrow$ P P $\rightarrow$ A\\ V $\rightarrow$ P P $\rightarrow$ V\\ T $\rightarrow$ P P $\rightarrow$ T\end{tabular}}\\
                            & paper (P): 14328       & P: 4231   &          \\
                            & term (T): 7723       & T: 50     &          \\
                            & venue (V): 20       & V: ---    &          \\ \bottomrule
\end{tabular}
}
\scriptsize
\vspace{-5pt}
\end{table}

\begin{figure}[!t] 
    \vspace{0pt}
	\centering
	\includegraphics[width=0.49\textwidth]{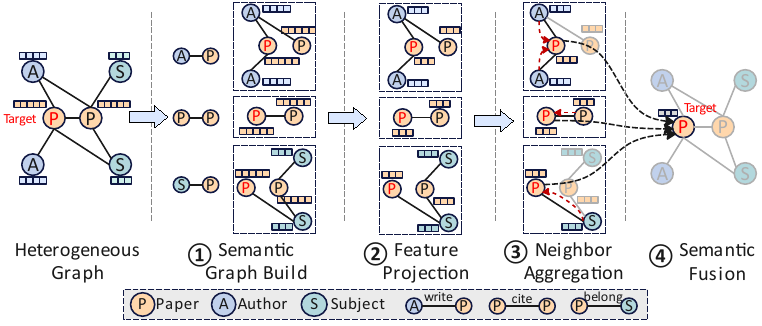}
	\vspace{-10pt}
	\caption{Illustration of HGNNs.}
	\label{fig:HGNN_work_flow}
	\vspace{-10pt}
\end{figure}

\noindent \textbf{HGNNs.} 
To generate the final embedding of each vertex, HGNNs recursively aggregate the feature vectors of its neighboring vertices in each semantic graph and fuse the aggregated results across all semantic graphs, as illustrated in Fig. \ref{fig:HGNN_work_flow}. The most prevalent HGNNs typically involve four stages.
The SGB stage constructs semantic graphs for the subsequent stages by partitioning the HetG into a group of semantic graphs based on relations or metapaths.
The FP stage projects the feature vector of each vertex in different types into the same dimensional space using an MLP within each semantic graph.
The NA stage utilizes an attention mechanism to perform a weighted sum aggregation of features from neighbors within each semantic graph.
The SF stage fuses the semantic information obtained from all semantic graphs, aiming to combine the results of the NA stage across different semantic graphs for each vertex.

\noindent \textbf{Differences between GNNs and HGNNs.}
Feature Projection: Vertices in HomoGs share the same vector space for joint feature projection; Vertices of different types in HetGs require separate feature projection.
Aggregation: GNNs only perform neighbor aggregation; HGNNs use neighbor aggregation and semantic fusion.



\section{Motivation}

This section quantitatively analyzes the buffer thrashing issue in HGNN acceleration to provide motivation for GDR-HGNN design. 

This issue significantly impedes the efficiency of HGNN acceleration, especially considering that the NA stage dominates HGNNs, which constituting up to 74\% of the total inference time~\cite{understand_HGNN}. 
It arises from the fact that the NA stage aggregates the features of neighboring vertices based on the irregular topology of semantic graphs, leading to irregular memory accesses to features.
To delve deeper into this matter, we perform quantitative analysis using the NVIDIA T4 GPU and the SOTA HGNN accelerator HiHGNN~\cite{HiHGNN}.

\textbf{Quantitative Analysis on T4 GPU.} 
We conduct a quantitative experiment for the NA stage of RGCN~\cite{RGCN} model using a state-of-the-art framework, DGL~\cite{DGL}, running on an NVIDIA T4 GPU.
The experimental results reveal that the L2 cache hit ratio in the processing of IMDB and DBLP is lower, reaching 30.1\% and 17.5\%, respectively. This reveals that a significant number of vertex features experience frequent replacements, contributing to the buffer thrashing issue and impairing overall performance.

\begin{figure}[!h] 
	\vspace{-5pt}
	\centering
	\includegraphics[width=0.48\textwidth]{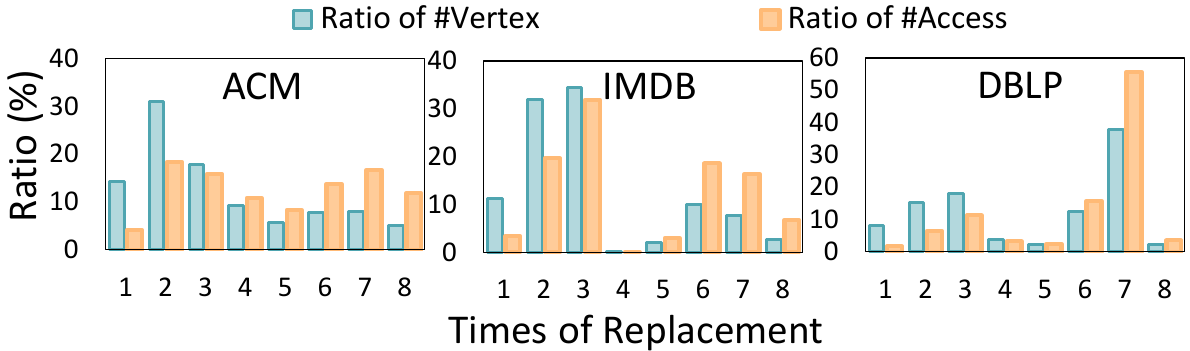}
	\vspace{-10pt}
	\caption{Analysis on HiHGNN with RGCN Model: Replacement Times of Vertices' Features during NA Stage.}
	\label{fig:motivation}
	\vspace{-10pt}
\end{figure}

\textbf{Quantitative Analysis on HiHGNN.} Fig.~\ref{fig:motivation} gives statistics on the replacement times of vertex features from the buffer during the NA stage. 
The numbers on the horizontal axis represent the replacement times of vertices' features, while ``Ratio of \#Vertex'' represents the ratio of the number of vertices with specific replacement times to the total number of vertices. Similarly, ``Ratio of \#Access'' denotes the ratio of number DRAM accesses conducted by vertices with specific replacement times to the total number of DRAM accesses. The results indicate that a considerable number of vertex features undergo frequent replacements, contributing to the buffer thrashing issue and resulting in a substantial number of redundant DRAM accesses. This excessive data movement significantly hinders overall performance.
It's worth noting the varying degrees of buffer thrashing across the three datasets, attributed to their different graph sizes. The DBLP dataset exhibits the most pronounced occurrence, primarily due to its significantly larger number of vertices compared to the other datasets.


\section{Design of GDR-HGNN} 

In this section, we initially identify an opportunity to tackle the buffer thrashing issue. 
To capitalize on this opportunity, we propose a graph restructuring method to enhance data locality in HGNN acceleration. 
Ultimately, we design GDR-HGNN, an HGNN accelerator frontend that incorporates this graph restructuring method.


\subsection{Opportunity to Address Buffer Thrashing}~\label{sec:opportunity}

To address the buffer thrashing issue in HGNN acceleration, an effective idea is to identify a group of incoming vertices that share the same neighboring vertices and aggregate their neighboring vertices during the same time frame. 
This exploitation of data locality can alleviate the irregular accesses to DRAM, resulting in a reduction of on-chip data replacement.

As mentioned in Section~\ref{sec:background}, HGNNs do not directly operate on the original HetG. 
Instead, the original HetG is processed to build several directed and bipartite semantic graphs~\cite{OGB}.
A bipartite graph can produce a maximal matching, namely largest number of edges that owning unique vertices. Based on the maximal matching, a minimal subset of vertices can be generated, within which all edges of the graph possess at least one endpoint.
This subset of vertices is termed the graph backbone in this paper, and it can facilitate the classification of all vertices into four distinct parts:
\ding{182} $Src_{in}$: Source vertices included in the backbone. \ding{183} $Src_{out}$: Source vertices excluded from the backbone. \ding{184} $Dst_{in}$: Destination vertices included in the backbone. \ding{185} $Dst_{out}$: Destination vertices excluded from the backbone.

By definition, there is no such edge whose endpoints are both outside the graph backbone, which derives the non-connectivity between $Src_{out}$ and $Dst_{out}$, and further a partition of the original graph into three distinct subgraphs.
These subgraphs demonstrate specific characteristics: all vertices connect to $Src_{out}$ belong to $Dst_{in}$, while all vertices connect to $Dst_{out}$ belong to $Src_{in}$. 

By eliminating irrelevant vertices from each subgraph, the processing of those shared neighboring vertices within the same timeframe, optimizing buffer utilization and mitigating the issue of buffer thrashing.


\subsection{Graph Restructuring Method}\label{sec:graph_restructure_alg}

To harvest the above opportunity, we propose a graph restructuring method to reshape input semantic graphs, leveraging the inherent properties of bipartite graphs to improve data locality.

\begin{figure}[!h] 
	\vspace{-10pt}
	\centering
	\includegraphics[width=0.48\textwidth]{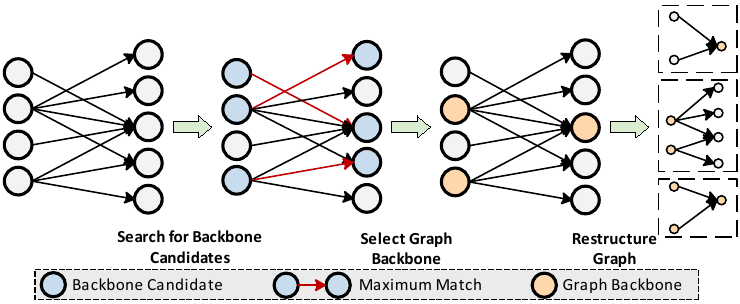}
	\vspace{-10pt}
	\caption{Toy Example of Graph Restructuring Method.}
	\label{fig:toy_example}
	\vspace{-10pt}
\end{figure}

Fig.~\ref{fig:toy_example} offers a toy example to elucidate the workflow of the graph restructuring method. 
This method unfolds in two stages: graph decoupling and graph recoupling. In the former stage, the maximum matching algorithm is employed to identify graph backbone candidates, effectively decoupling the semantic graph into a distinct edge group. In the latter stage, the backbone is selected from this discrete edge group to reassemble the graph into three subgraphs, each characterized by a robust community structure.

\textbf{Graph Decoupling.}
The graph decoupling focuses on discovering the maximum matching within semantic graphs to identify graph backbone candidates, as illustrated in Algorithm~\ref{alg:decouple_algorithm}. Essentially, it draws inspiration from the Hungarian Algorithm~\cite{Hungarian_algorithm}. To optimize its execution, we design customized hardware by leveraging FIFO and hash table functionalities, as detailed in Section~\ref{sec:architecture}.

This algorithm begins by initializing all vertices without a match (line 2), subsequently scanning each vertex $u$ to identify its neighbor $v$ and commence the search.
Initially, for each neighbor $v$ of $u$, $u$ is added to the $Matchin\_FIFO$[$v$] (line 12) for temporary storage. If $v$ is unmatched, the algorithm records vertices $u$ and $v$ as a matched pair and frees the previous match associated with $u$ through iterative steps (lines 14-18).
In instances where all of $u$'s neighbors are matched, the vertices matched to $u$'s neighbors are pushed into $Search\_{List}$ (lines 22-26) to find another match, leaving the match for $u$.
After searching for the maximum match, the final matches are stored in $Match\_Pair$. Additionally, the matching vertices are recorded as graph backbone candidates, termed as $M$.

\begin{algorithm}[!htbp]
    \SetAlgoLined
    \small
    \caption{\textbf{Graph Decoupling}}
    \label{alg:decouple_algorithm}
    \KwIn{$G^\mathcal{P}$: The input semantic graph;}
    \KwOut{$Match\_Pair$: Backbone candidate list;}
    $Match\_Pair$, $Search\_List$ = \{ \};\\
    Clear all $Matching\_FIFO$;\\
    \For{each vertex $n$ in $G^\mathcal{P}$}{
        \If{$Match\_Pair$[$n$] < 0}{
            Push $n$ to $Search\_List$;\\
            \While{$Search\_List$ is not empty}{
                Pop $u$ from $Search\_List$;\\
                \For{each neighbor $v$ of $u$}{
                    \If{$v$ is visited}{
                       continue;\\
                    }
                    Push $u$ to $Matching\_FIFO$[$v$];\\
                    \If{$Match\_Pair$[$v$] < 0}{
                        \While{$Match\_Pair$[$u$] > 0}{
                            $Matching\_FIFO$[$Match\_Pair$[$u$]].pop();\\
                            Change $Match\_Pair$;\\
                            $u$ = $Match\_Pair$[$Match\_Pair$[$u$]];\\
                        }
                        break while;\\
                    }
                }
                \If{$Match\_Pair$[$v$] < 0}{
                    \For{each matched neighbor $u$ of $v$}{
                        Push $Match\_Pair$[$u$] to $Search\_List$;\\
                    }
                }
            }
        }
    }
    return $Match\_Pair$;\\
\end{algorithm}

\textbf{Graph Recouping.}
The graph recoupling aims to select the graph backbone from the candidates and generate subgraphs, as shown in Algorithm~\ref{alg:recouple_algorithm}.


Initially, the algorithm initiates backbone selection. It commences by exploring all matched vertices categorized by source and destination. For each matched source vertex, the algorithm identifies its unmatched neighbors and put them to $Dst_{out}$ while put itself to $Src_{in}$. If all its neighbors are matched, the algorithm will just skip to the next matched source vertex. Also, all matched destination vertices will be examined after the matched sources vertices for the same procedure. After all, all left vertices are then classified to $Src_{out}$ or $Dst_{out}$, according to whether they belong to $V_{src}$ or $V_{dst}$.
Once the backbone is selected, subgraphs are generated for the subsequent execution.

\begin{algorithm}[!h]
    \SetAlgoLined
    \small
    \caption{\textbf{Graph Recoupling}}
    \label{alg:recouple_algorithm}
    \KwIn{$G^\mathcal{P}$: The input graph;  $M$: Backbone candidate vertex set;}
    \KwOut{$G_{s_1}^{\mathcal{P}}$, $G_{s_2}^{\mathcal{P}}$, $G_{s_3}^{\mathcal{P}}$: The subgraphs generated from original semantic graph.}
    $Src_{in}, Src_{out}, Dst_{in}, Dst_{out}$ = \{ \};\\
    $S\leftarrow V_{src}\cap M, \overline{S}\leftarrow V_{src}\setminus M, T\leftarrow V_{dst}\cap M, \overline{T}\leftarrow V_{dst}\setminus M$; \\
    \For{each $v$ in $S$}{
        $X_v \leftarrow N(v)\cap \overline{T}$; \\
        \If{$X_v$ is not $\varnothing$}{
            Push $v$ to $Src_{in}$;\\
            Push $X_v$ to $Dst_{out}$;\\
        }
    }
    \For{each $u$ in $T$}{
        $X_u \leftarrow N(u)\cap \overline{S}$; \\
        \If{$X_u$ is not $\varnothing$}{
            Push $u$ to $Dst_{in}$;\\
            Push $X_u$ to $Src_{out}$;\\
        }
    }
    Push the other source vertices to $Src_{out}$;\\
    Push the other destination vertices to $Dst_{out}$;\\
    $G_{s_1}^{\mathcal{P}}$, $G_{s_2}^{\mathcal{P}}$, $G_{s_3}^{\mathcal{P}}$ = \textbf{GenerateGraph}($Src_{in}$, $Src_{out}$, $Dst_{in}$, $Dst_{out}$);\\
    return $G_{s_1}^{\mathcal{P}}$, $G_{s_2}^{\mathcal{P}}$, $G_{s_3}^{\mathcal{P}}$;
\end{algorithm}



\subsection{Hardware Implementation}\label{sec:architecture}





In this section, we detail the algorithm mapping to the hardware and the microarchitecture of GDR-HGNN.

\begin{figure}[!hptb] 
	\vspace{-5pt}
	\centering
	\includegraphics[width=0.45\textwidth]{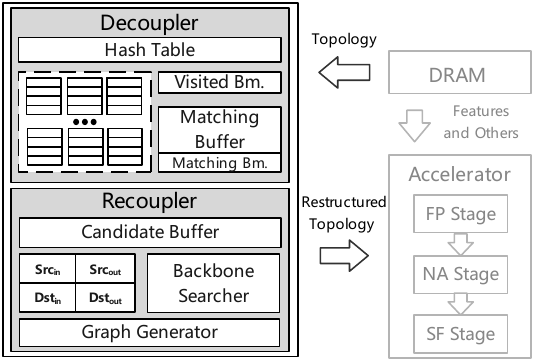}
	\vspace{-10pt}
	\caption{Design and Workflow Overview of GDR-HGNN.}
	\label{fig:architecture}
	\vspace{-10pt}
\end{figure}

Fig.~\ref{fig:architecture} offers an overview of the GDR-HGNN architecture, which primarily comprises two modules: Decoupler and Recoupler. 
The Decoupler undertakes graph decoupling and is constructed with components of the Hash Table, FIFOs, bitmaps (Bm.), and buffers. 
On the other hand, the Recoupler is tasked with executing graph recoupling and consists of the Backbone Searcher, a collection of FIFOs, and the Graph Generator.


\begin{figure}[!h] 
	\centering
	\includegraphics[width=0.35\textwidth]{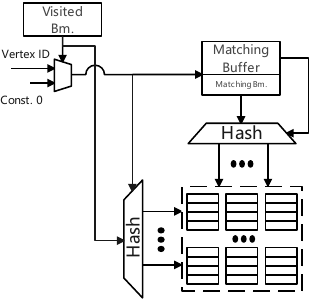}
	\vspace{-10pt}
	\caption{Micro-architecture of Decoupler.}
	\label{fig:decoupler_micro_architecture}
	\vspace{-10pt}
\end{figure}

\textbf{Workflow of Decoupler.} In each execution epoch, the topology of original semantic graph is received and passed on to the hash table for FIFO allocation.
The FIFOs, organized in a set-associative manner, store matched pairs and waiting list allocated to specific vertices. 
As illustrated in Fig.~\ref{fig:decoupler_micro_architecture}, during each cycle, source vertices are dispatched to their respective FIFOs, automatically triggering a pop operation for FIFOs if the match condition changes. 
The Matching Buffer stores replaced FIFO data. Upon identifying all matching edges, the resulting backbone candidates are stored in each FIFO and sent to the Candidate Buffer.

\begin{figure}[!h] 
	\centering
	\includegraphics[width=0.38\textwidth]{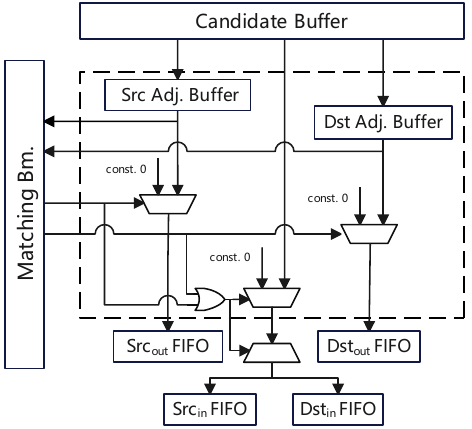}
	\vspace{-10pt}
	\caption{Micro-architecture of Recoupler.}
	\label{fig:recoupler_micro_architecture}
	\vspace{-10pt}
\end{figure}

\textbf{Workflow of Recoupler.} The identified candidates are then forwarded to the Backbone Searcher, as highlighted in Fig.~\ref{fig:recoupler_micro_architecture}. 
During this stage, the Candidate Buffer transmits the backbone candidates to the Backbone Searcher to identify the graph backbone, following Algorithm~\ref{alg:recouple_algorithm}.
Initially, each candidate is directed to the adjacency list buffers including the Src Adj. Buffer and Dst Adj. Buffer to obtain their respective neighbors. Subsequently, all obtained neighbors are checked in the Matching Bm. If any neighbors are not found in the Matching Bm., the candidate is sent to either $Src_{in}$ or $Dst_{in}$ FIFOs, depending on its origin, and the corresponding neighbors are dispatched to the $Src_{out}$ or $Dst_{out}$ FIFOs.
The Graph Generator creates the subgraphs from these four designated buffers and forwards them for subsequent HGNN execution.

\textbf{Cooperation with HGNN Accelerators.} 
To illustrate how to collaborate with HGNN accelerators, we present a case study using the SOTA accelerator HiHGNN. It's important to note that the graph restructuring method and its scheduling unit design can be integrated with other future HGNN accelerators for enhanced performance. 
Additionally, this method can be applied to subgraphs to generate smaller sub-subgraphs, thereby exploiting data locality in a smaller on-chip buffer.
HiHGNN~\cite{HiHGNN} employs a hybrid architecture that includes a systolic array module for matrix-vector multiplication and a SIMD module to perform element-wise operations during HGNN execution, covering FP, NA, and SF stages.

By establishing a dataflow between GDR-HGNN and HiHGNN, efficient HGNN acceleration can be achieved through the pipelining of their execution. They operate concurrently and share the memory controller to manage data interactions between on-chip buffers and high-bandwidth memory (HBM).
The process begins with the input of a semantic graph into GDR-HGNN, where the subgraphs, extracted from the FIFOs including $Src_{out}$ and $Dst_{in}$, $Src_{in}$ and $Dst_{in}$, and $Src_{in}$ and $Dst_{out}$, are transmitted to HiHGNN. Once the graph backbone is identified. Meanwhile, GDR-HGNN continuously receives and restructures the next semantic graph.

\section{Evaluation}
In this section, we compare GDR-HGNN to baselines and provide the optimization analysis in detail.
We first describe our experimental setup in Section~\ref{sec:exp_setup}. 
Then, we demonstrate the advantages with the assistance of GDR-HGNN, comparing against a SOTA software framework operating on both NVIDIA T4 GPU and A100 GPU, as well as a SOTA HGNN accelerator HiHGNN in Section~\ref{sec:exp_results}.


\subsection{Experiment Setup}\label{sec:exp_setup}

\noindent\textbf{Evaluation Methodology.}
The performance metrics of GDR-HGNN are evaluated using the following tools.

\textit{Architecture Simulator.}
We implement GDR-HGNN in a cycle-level accurate simulator to measure execution time in the number of cycles. We also design a detailed cycle-accurate on-chip memory model and integrate the Ramulator~\cite{ramulator} for FIFOs, buffers and memory simulation. This simulator models the microarchitectural behaviors of each module and the hardware datapath.

\textit{CAD Tools.}
We implement an RTL version of each hardware module and synthesize it to evaluate the area, energy consumption, and latency. We use the Synopsys Design Compiler with the TSMC 12 \textit{nm} standard VT library for the synthesis and estimate the power consumption using Synopsys PrimeTime PX.

\textit{Memory Measurements.}
We estimate the buffer area, energy consumption, and access latency using Cacti~\cite{CACTI}. We use four different scaling factors to convert them to 12 nm technology.
The access latency and energy of HBM 1.0 are simulated by the Ramulator and estimated with 7 pJ/bit as HiHGNN, respectively.

\noindent\textbf{Datasets and Models.}
We conduct the experiments on three different models including RGCN~\cite{RGCN}, RGAT~\cite{RGAT} and Simple-HGN~\cite{Simple-HGN}, using three datasets, ACM, DBLP, and IMDB, commonly used in HGNN research community~\cite{DGL, understand_HGNN, HiHGNN}. The implementation of all models follows the specifications outlined in HiHGNN~\cite{HiHGNN}.

\noindent\textbf{Baseline Platforms.} We integrate GDR-HGNN into HiHGNN, creating the combined system HiHGNN+GDR-HGNN. We compare it with a state-of-the-art HGNN framework, DGL 1.0.2~\cite{DGL}, running on an NVIDIA T4 GPU, an NVIDIA A100 GPU, and HiHGNN.
Table~\ref{tb:baselines} lists the configurations for HiHGNN and GDR-HGNN.






\begin{table}[!t]
\centering
\caption{Platform Details of HiHGNN and GDR-HGNN.}\label{tb:baselines}
\vspace{-10pt}
\renewcommand\arraystretch{1.2}
\resizebox{0.47\textwidth}{!}{
\begin{tabular}{ccc}
\toprule
 & \textbf{HiHGNN} & \textbf{GDR-HGNN} \\ \hline
\begin{tabular}[c]{@{}c@{}}Peak Performance\end{tabular} & 

\multicolumn{1}{c}{\begin{tabular}[c]{@{}c@{}}16.38 TFLOPS, 1.0 GHz\end{tabular}} & --- \\ \hline
On-chip Buffer & \begin{tabular}[c]{@{}c@{}}2.44 MB (FP-Buf),\\ 14.52 MB (NA-Buf),\\ 0.12 MB (SA-Buf),\\ 0.38 MB (Att-Buf)\end{tabular} & \begin{tabular}[c]{@{}c@{}}8 KB FIFOs,\\ 160 KB Matching Buffer,\\ 160 KB Candidate Buffer,\\ 320 KB Adj. List Buffer\end{tabular} \\ \hline
Off-chip Memory &
\multicolumn{1}{c}{\begin{tabular}[c]{@{}c@{}}512 GB/s, HBM 1.0\end{tabular}} & --- \\ \hline
\end{tabular}
}
\end{table}

\subsection{Evaluation Results}\label{sec:exp_results}

\begin{figure}[!h] 
	\centering
	\includegraphics[width=0.48\textwidth]{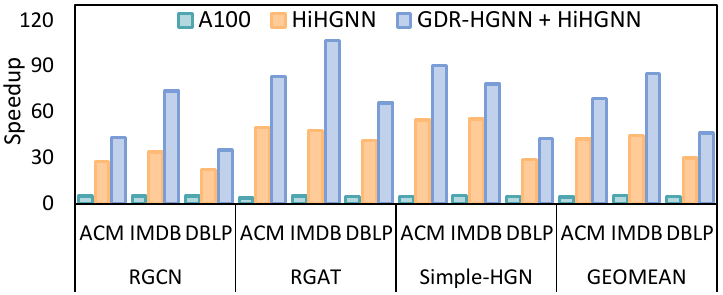}
	\vspace{-10pt}
	\caption{Speedup to T4 GPU.}
	\label{fig:speedup}
	\vspace{-10pt}
\end{figure}

\noindent\textbf{Speedup.}
Fig.~\ref{fig:speedup} shows the speedup of A100 GPU, HiHGNN, and HiHGNN+GDR-HGNN to T4 GPU. The last set of bars, labeled as GEOMEAN, indicates the geometric mean across all HGNN models. 
HiHGNN+GDR-HGNN achieves an average speedup of 68.8$\times$, 14.6$\times$ and 1.78$\times$ compared to T4 GPU, A100 GPU and HiHGNN, respectively.
The performance improvement stems from two primary aspects. Firstly, the graph restructuring method facilitates the transformation of graph structures, evolving from a structure that generates a large number of random accesses to an organized form showcasing community locality. This restructure helps reduce buffer replacements, thereby enhancing overall performance.
Secondly, the pipeline between Decoupler, Recoupler and accelerator ensures uninterrupted utilization of intermediate data. This design further reduces buffer replacements, leading to an overall performance boost.


\begin{figure}[!h] 
	\centering
	\includegraphics[width=0.48\textwidth]{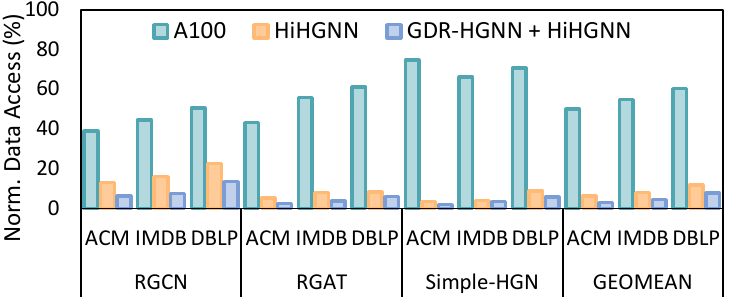}
	\vspace{-15pt}
	\caption{Normalized DRAM Access to T4 GPU.}
	\label{fig:DRAM_access}
	\vspace{-10pt}
\end{figure}

\noindent\textbf{Number of DRAM Accesses.}
To analyze the source of performance improvement, Fig.~\ref{fig:DRAM_access} presents the normalized DRAM access to T4 GPU. HiHGNN+GDR-HGNN accesses only 4.8\%, 8.7\%, and 57.1\% compared to T4 GPU, A100 GPU, and HiHGNN, respectively. This result confirms that with the assistance of GDR-HGNN, HiHGNN significantly reduces the number of DRAM accesses, validating the source of speedup.

\begin{figure}[!h] 
	\centering
	\includegraphics[width=0.48\textwidth]{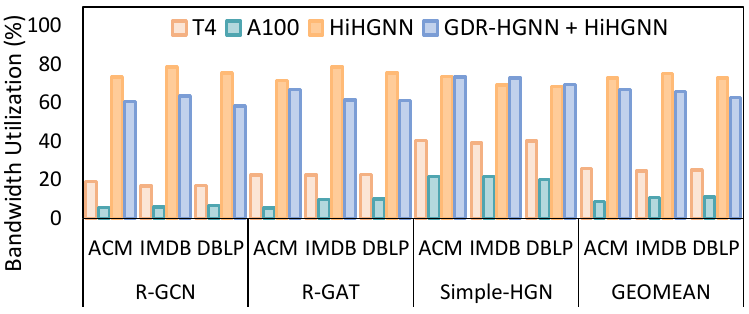}
	\vspace{-10pt}
	\caption{DRAM Bandwidth Utilization.}
	\label{fig:bandwidth_utilization}
	\vspace{-8pt}
\end{figure}

\noindent\textbf{Utilization of DRAM Bandwidth.}
Fig.~\ref{fig:bandwidth_utilization} shows the utilization of the DRAM bandwidth of GDR-HGNN+HiHGNN and the baselines. 
GDR-HGNN+HiHGNN demonstrates 2.58$\times$ and 6.35$\times$ improvement on average in the utilization of DRAM bandwidth compared with T4 GPU and A100 GPU, respectively.
In contrast to HiHGNN, HiHGNN+GDR-HGNN notably diminishes DRAM accesses using the graph restructuring method. However, this improvement comes with a marginal trade-off affecting overall bandwidth utilization, primarily due to increased strain on compute resources.

\begin{figure}[!h] 
	\centering
	\includegraphics[width=0.49\textwidth]{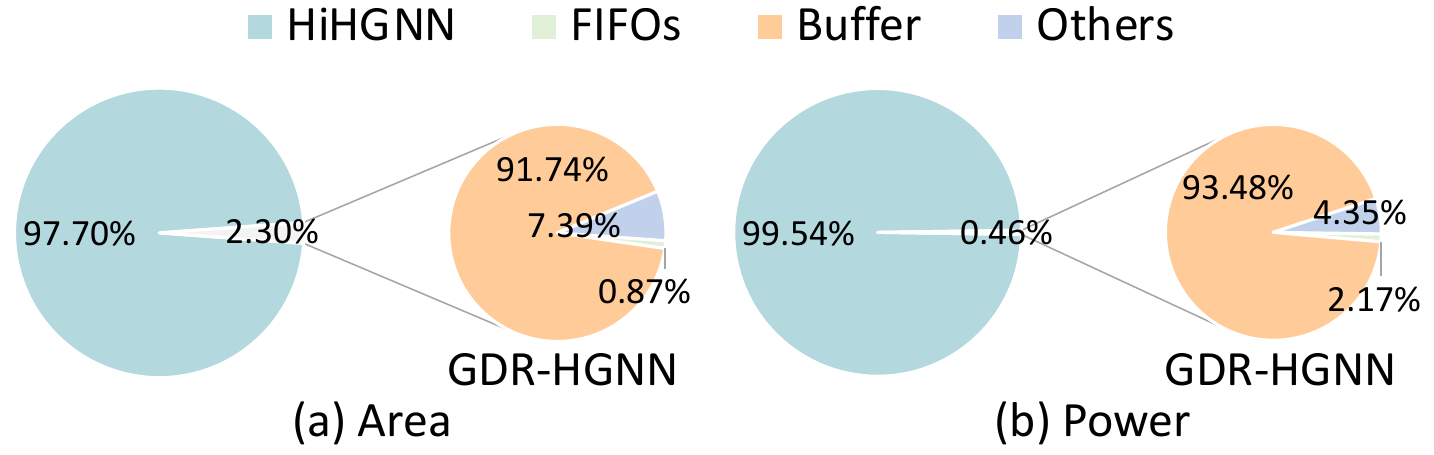}
	\vspace{-10pt}
	\caption{Area and Power of HiHGNN and GDR-HGNN.}
	\label{fig:power_area_overhead}
	\vspace{-10pt}
\end{figure}

\noindent\textbf{Area and Power Overhead.}
Fig.~\ref{fig:power_area_overhead} displays the area and power characteristics of HiHGNN and GDR-HGNN. The results indicate that GDR-HGNN accounts for only 2.30\% (i.e., 0.50 $mm^2$) and 0.46\%  (i.e., 55.6 $mW$) of the total area and power when combined with HiHGNN under TSMC 12 $nm$ technology. This validates that the overhead of the graph restructuring method can be disregarded. The primary overhead of GDR-HGNN originates from buffers used to store edge and vertex indices for the restructure.



\section{Related Work}

Given the remarkable learning capacity of GNNs with graph data, GNN accelerators have garnered significant attention from the architecture community~\cite{HyGCN,igcn,FlowGNN,rrgat}. In terms of data locality exploitation, I-GCN~\cite{igcn} introduces a potent method called ``islandization'' to enhance data locality in GNNs. This method identifies clusters of vertices with strong internal connections but weak external connections.
However, this method is not suitable for directed bipartite graphs, as the properties of such graphs cause the employed strategy to degrade into a process focused solely on finding the vertex with the largest degree.

Previous efforts~\cite{HiHGNN,MetaNMP} have proposed several accelerators for HGNN acceleration.
HiHGNN~\cite{HiHGNN} strategically schedules the execution order of semantic graphs based on their similarity to exploit data reusability across different semantic graphs.
MetaNMP~\cite{MetaNMP} pioneers DIMM-based near-memory processing for HGNNs. It employs a cartesian-like product paradigm to dynamically generate metapath instances and aggregate vertex features from the starting vertex along these metapaths.
In contrast, our work focuses on alleviating buffer thrashing issues by leveraging the opportunity presented by the semantic graphs in HGNNs.


\section{Conclusion}

This work identifies an opportunity for data locality exploitation via an in-depth analysis of the unique characteristics of HetGs. It introduces GDR-HGNN, designed to leverage this opportunity to address the buffer thrashing issue during HGNN execution. Experimental results show that GDR-HGNN outperforms state-of-the-art efforts and substantially reduces DRAM accesses.
 



\section*{Acknowledgment}
This work was supported by National Key Research and Development Program (Grant No. 2022YFB4501404), the National Natural Science Foundation of China (Grant No. 62202451), CAS Project for Young Scientists in Basic Research (Grant No. YSBR-029), and CAS Project for Youth Innovation Promotion Association.

\bibliographystyle{ACM-Reference-Format}
\bibliography{ref}

\end{document}